\documentclass[a4paper,conference]{IEEEtran}
\usepackage{algorithm}
\usepackage{algpseudocode}
\usepackage{amsmath}
\usepackage{amssymb}
\usepackage{mathpazo}
\usepackage{booktabs}
\usepackage{url}

\ifCLASSINFOpdf
  \usepackage[pdftex]{graphicx}
\else
\fi
\ifCLASSOPTIONcompsoc
  \usepackage[caption=false,font=normalsize,labelfont=sf,textfont=sf]{subfig}
\else
  \usepackage[caption=false,font=footnotesize]{subfig}
\fi
\begin{document}
%

\title{Online Assessment Misconduct Detection using Internet Protocol and Behavioural Classification}

\author{\IEEEauthorblockN{Leslie Ching Ow Tiong}
\IEEEauthorblockA{Computational Science Research Center\\
Korea Institute of Science and Technology\\
Seoul, Republic of Korea\\
Email: tiongleslie@kist.re.kr}
\and
\IEEEauthorblockN{HeeJeong Jasmine Lee}
\IEEEauthorblockA{Pierson College\\
PyeongTaek University\\
Gyeonggi-Do, Republic of Korea\\
Email: hjlee@ptu.ac.kr}
\and
\IEEEauthorblockN{Kai Li Lim}
\IEEEauthorblockA{Dow Centre\\
The University of Queensland\\
Brisbane, Australia\\
Email: kaili.lim@uq.edu.au}
}


%


\maketitle

\begin{abstract}
With the recent prevalence of remote education, academic assessments are often conducted online, leading to further concerns surrounding assessment misconducts. This paper investigates the potentials of online assessment misconduct (e-cheating) and proposes practical countermeasures against them. The mechanism for detecting the practices of online cheating is presented in the form of an e-cheating intelligent agent, comprising of an internet protocol (IP) detector and a behavioural monitor. The IP detector is an auxiliary detector which assigns randomised and unique assessment sets as an early procedure to reduce potential misconducts. The behavioural monitor scans for irregularities in assessment responses from the candidates, further reducing any misconduct attempts. This is highlighted through the proposal of the DenseLSTM using a deep learning approach. Additionally, a new PT Behavioural Database is presented and made publicly available. Experiments conducted on this dataset confirm the effectiveness of the DenseLSTM, resulting in classification accuracies of up to 90.7\%.
\end{abstract}


%
\IEEEpeerreviewmaketitle

\section{Introduction}
The proliferation of online learning delivery methods has resulted in wide adoptions across all levels of education that are further catalysed by the recent pandemic \cite{Almaiah2020, Adedoyin2020}. Still, the impersonal nature of these methods raises concerns about the potential risks of academic dishonesty, particularly when assessments are remotely conducted, as they often lack the disciplinary procedures and invigilation that are conventionally used in in-person assessment centres \cite{Sarrayrih2013, Jalali2017, Raud2019}. With a rapidly increasing uptake in online learning, existing literature indicates a prevalence of assessment misconducts \cite{Rowe2004, Moten2013, Chuang2017, Clark2020, Elsalem2021}. In other words, while online learning enables these opportunities to be more accessible, its credibility may be compromised if these issues remain unresolved.

Assessment misconducts are a common occurrence across the world. To mitigate these problems, early work done by \cite{Roger2006} suggested the installation of proctoring security software, which enables invigilators to continuously monitor the computer screen of candidates through an instructor control view. 
Similarly, \cite{Cluskey2011} proposed an approach to reduce potential misconducts by considering the criteria that were contributory to the assessment, such as assessment duration and time allocation for each assessment question. However, this approach has limitations given that it required the students to use a pre-specified Internet browser to access the assessment, with assessors varying the question sets upon each delivery.

With the advancement of wireless telecommunications, many works aim to strengthen the security of online assessment protocols to counter the security breaches that could occur. A review by \cite{Ullah2016} and \cite{Ullah2018} highlighted the significance of innovations to enable techniques such as anomaly detection to address online assessment misconducts (e-cheating). Candidates were remotely monitored at an online assessment, where their behaviours were simultaneously assessed using randomised questions presented through a dynamic profile. Results showed that candidates who impersonate shared most information by collaborating through online messengers or a mobile device, to which the response times were significantly different to those who were more honest.

The identification of candidates is often verified through their biometrics. In contrast to in-person assessments, online assessments do not necessarily employ proctors or invigilators and can be held across different uncontrolled remote environments. For instance, \cite{Hu2018} and \cite{Garg2020} proposed using face recognition to monitor the candidates' behaviour during the assessments, identifying the relevant norms between the image of a candidate's face and their corresponding pose. However, such an approach requires the candidates to have webcams and a robust and stable network connection, which may be prohibitive to rural candidates. 

This paper aims to address the current problems of e-cheating through artificial intelligence (AI) techniques using a network internet protocol (IP) detector and a deep learning-based behavioural monitor. The outcomes of this research offer avenues for further improving the proposed online assessment system. The main contributions of this paper are given as follows:
\begin{itemize}
\item The proposal of an e-cheating intelligent agent. This is based on a relationship model which detects misconduct using AI techniques. An IP detector and a behavioural monitor were designed, with the behavioural monitor using a densely connected long short-term memory (LSTM) network known as DenseLSTM; the IP detector is implemented as an auxiliary approach. This combination of techniques has never been attempted in literature to the best of our knowledge.
\item The curation and presentation of a new database for this study. This is made available in \cite{ourdb2020}. The contained records were gathered across several online assessments conducted in highly unregulated environments (e.g., without e-cheating mitigation). This database includes training and testing schemes for performance analysis and evaluation.
\end{itemize}

This paper is organised as follows. Section \ref{sec::sec2} presents our proposed framework with the new database detailed in Section \ref{sec::sec3}. Section \ref{sec::sec4} describes the experimental setup, results and discussions. Finally, the conclusions are summarised in the last section.

\section{E-cheating Intelligent Agent}
\label{sec::sec2}
\begin{figure}[!t]
\centering
\includegraphics[width=0.40\textwidth]{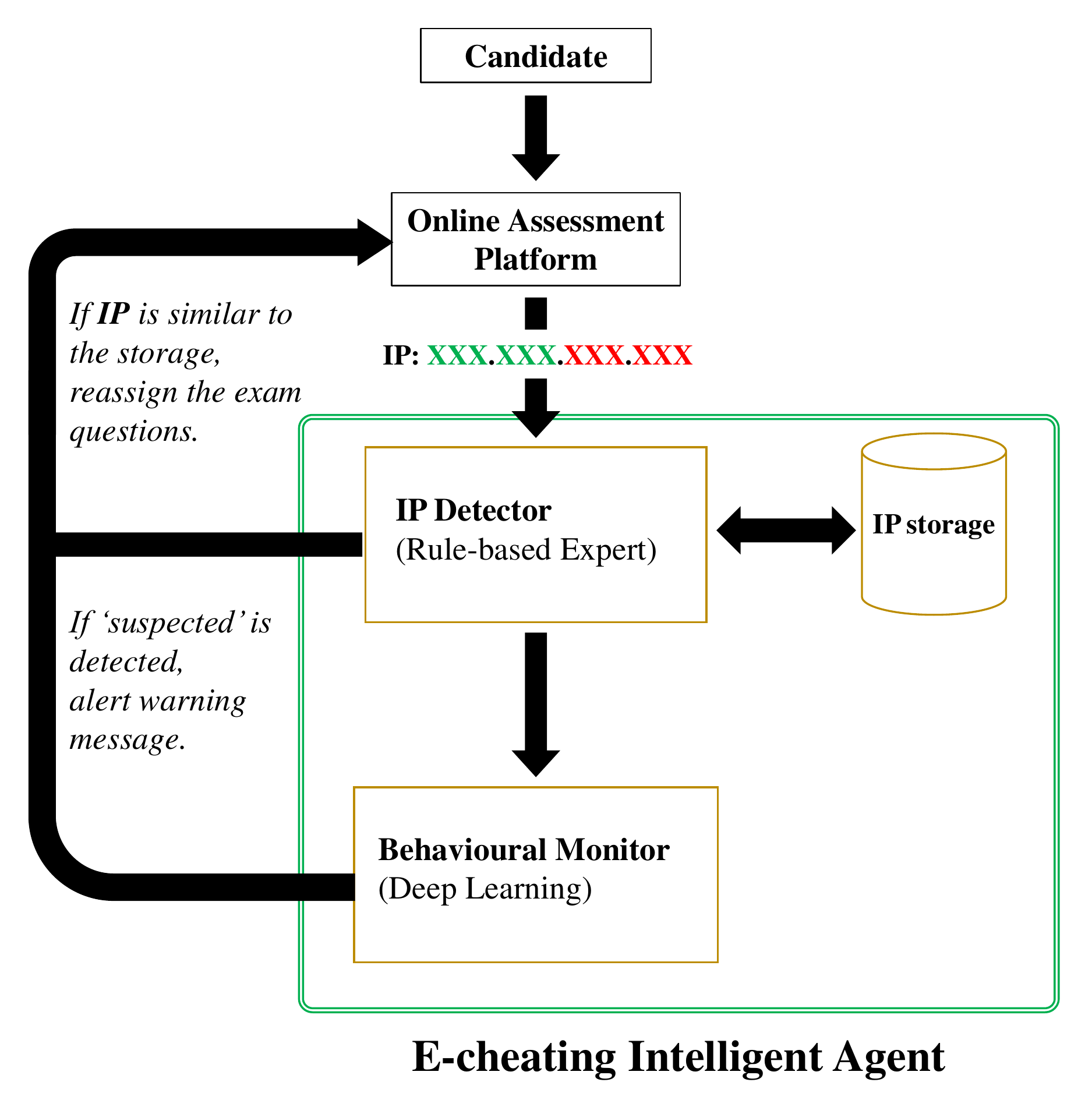}
\vspace{-10pt}
\caption{Proposed framework for the e-cheating intelligent agent system.}
\vspace{-1.0em}
\label{fig::fig1}
\end{figure}

This agent is designed to detect any potential misconduct, consisting of two main components: a network IP detector (described in Section \ref{subsec::subsec21}) and a behavioural monitor (described in Section \ref{subsec::subsec22}). Fig. \ref{fig::fig1} illustrates the architecture of this proposed system. This framework describes the system's process of identifying suspicious or `suspected' behaviours and preventing misconduct at online assessments. Note that the IP subnets with `red' refer to the `suspicious' cases, and IP subnets with `green' refer to the `normal' cases.

\subsection{Network IP Detector}
\label{subsec::subsec21}
Referring to Fig. \ref{fig::fig1}, we propose an IP detector module as an additional authentication measure that includes IP address monitoring. This detector considers that most Dynamic Host Configuration Protocol (DHCP) servers in household network routers allocate dynamic IP addresses---numerical labels (subnets) that identify each client device connected to a computer network. Any change in this IP pattern would enable the system to issue an alert, relating to scenarios like a candidate changing their computer device or their initial location. 

In this method, assessment questions are given as sets (such as Set A, B, C, etc.). Note that randomisation only applies to the question sequences with their answering choices shuffled for each set of questions. New questions can also be added such as those of similar topics within the course. 
An assessment starts with the IP address verification as outlined in Fig. \ref{fig::fig1}. Each candidate is initially assigned a random assessment question set (e.g., Set A). The incoming candidate then inputs their platform credentials and the IP detector looks up their IP address from the IP storage. If that IP address was previously flagged in the IP list as `suspicious' (the IP address is detected by the agent and appeared as `red' in Fig \ref{fig::fig1}), the agent will then generate a different set of questions (e.g., Set B) to the candidate. Algorithm \ref{alg::alg1} summarises the entire process of the IP detection agent.

\begin{algorithm}
\caption{IP Detector}
\label{alg::alg1}
\small
\textbf{Input:} IP address from a new candidate $E$\newline
\textbf{Output:} Decision $D$\newline \newline
\textbf{//Initialisation}\newline
Let $\text{IP}_{DB}$ be the list of real-time IP of size $N$\newline \newline
\textbf{if} $E$ is not in $\text{IP}_{DB}$ \textbf{then}
\begin{algorithmic}
\State Add $E$ into $\text{IP}_{DB}$
\State \textbf{return} $D$ with random sets to $E$
\end{algorithmic}
\textbf{else}
\begin{algorithmic}
\State \textbf{for} $i \rightarrow$ 1 to $N$ \textbf{do}
\State $\;\;\;$ \textbf{if} $E$ is found in $\text{IP}_{DB}[i]$ \textbf{then}
\State $\;\;\;\;\;\;\;$ \textbf{return} $D$ with specific sets to $E$
\State $\;\;\;$ \textbf{end}
\State \textbf{end}
\end{algorithmic}
\textbf{end}
\end{algorithm}

\subsection{Behavioural Monitor}
\label{subsec::subsec22}
\begin{figure}[!t]
\centering
\includegraphics[width=0.43\textwidth]{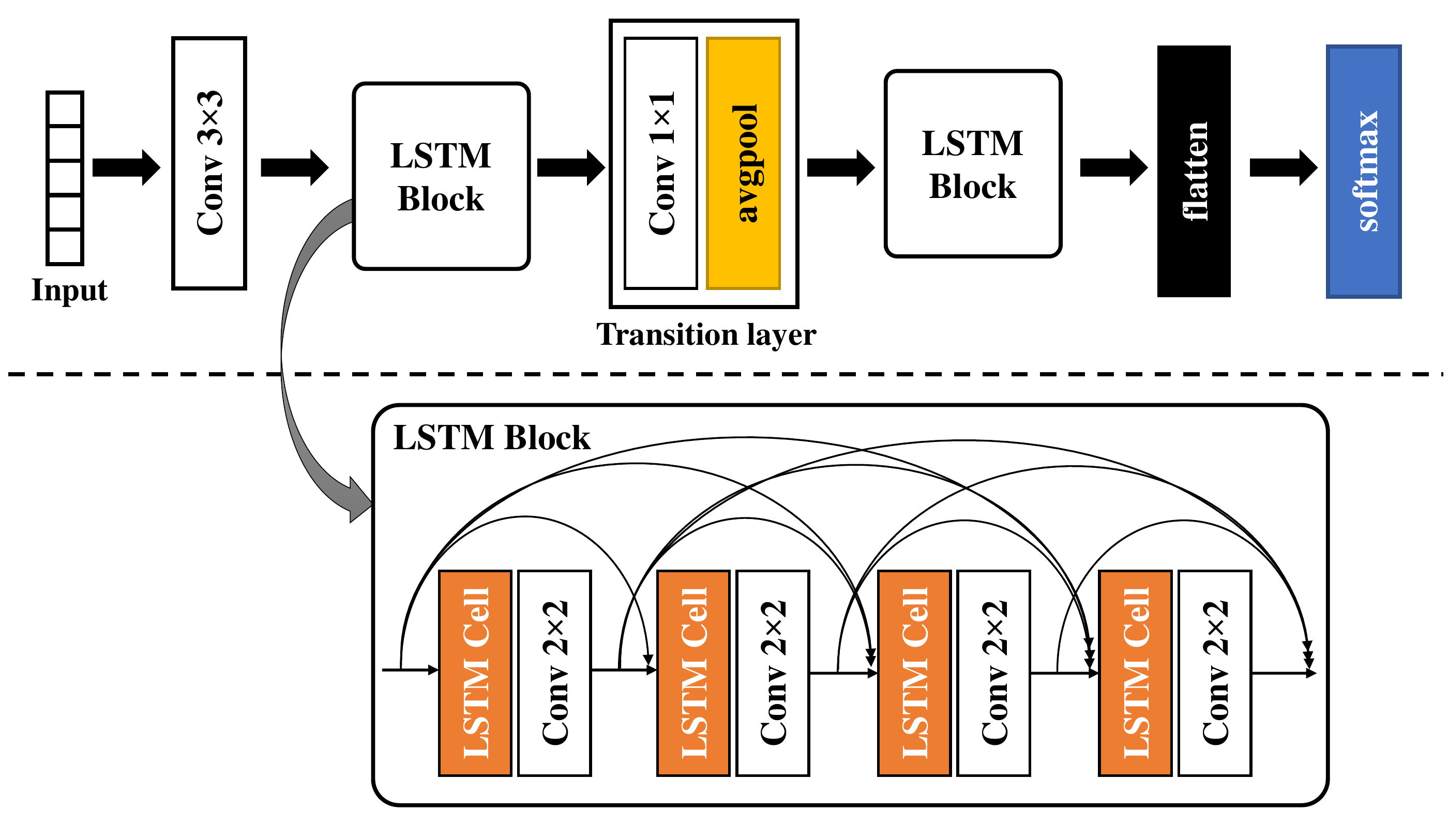}
\vspace{-10pt}
\caption{Architecture of the DenseLSTM.}
\vspace{-1.5em}
\label{fig::fig2}
\end{figure}

We devised a behavioural monitor module using a deep learning approach to monitor and analyse the behaviour of candidates. Referring to Fig. \ref{fig::fig2}, this agent would, when triggered, alert invigilators and immediately reassign the remaining questions with a new set. This trigger occurs when abnormal behaviour is detected in the candidate's answering pattern. The following subsections provide a more detailed explanation of this behavioural monitor.

Forming the cornerstone to this agent is DenseLSTM, a deep learning network which we have designed and modified from a long short-term memory (LSTM) network using  a densely connected approach. The LSTM network was introduced by \cite{Hochreiter1997}, and allows modellings with sequentially dependent data, such as time series, natural language processing and behavioural analysis. Using a densely connected approach allows better feature representations to predict behaviour abnormalities. 

Fig. \ref{fig::fig2} illustrates the architecture of DenseLSTM, which consists of a convolutional (\textit{conv}) layer, two LSTM blocks and a transition layer. Within the LSTM block, the LSTM cell layer determines the information to discard from the cell state. This layer could potentially and inadvertently omit useful information, we conceptualise a densely connected network (DenseNet) to retain information instead of the information handling method mentioned. The concept of DenseNet was originally proposed by \cite{Huang2019} for image classification. This network introduces direct connections from each layer to all subsequent layers, which improves information flow by creating a unique connectivity pattern. Each layer now has improved access to all preceding feature maps in a given block due to this dense connectivity, resulting in the provisioning of a ``collective knowledge'' within the network. Here, data features can be accessed from anywhere within the network, and unlike conventional network architectures, there is no need to replicate them across layers. When used in this study, this network is expected to collect useful information while maintaining low feature complexity, which can result in better classification performances.

The LSTM block $B$ is expressed in Equation \ref{eq::eq1} with $l$ in $H$ layers, comprising the LSTM cell, $conv$ layer, rectified linear unit (ReLU) and dropout layers.
\begin{equation}
B = H_{l}([x_{0}, x_{1}, x_{2}, \cdots, x_{l-1} ]),
\label{eq::eq1}
\end{equation}
where $x_{0}$ to  $x_{l-1}$ denotes feature outputs and $[\cdot]$ is defined as a concatenation operator. We set $l = 4$ in the first LSTM block and $l = 8$ in the second LSTM block. 

A transition layer is implemented in the first LSTM block that executes $1\times 1$ $conv$ and $avgpool$ operations, where $1\times 1$ $conv$ is defined as the filter size of the $conv$ layer that is $1\times 1$. This transition layer is used to control the complexity of the feature maps. Table \ref{tab::tab1} tabulates the architecture of this network.

In the training stage, we implement a softmax cross-entropy $\mathcal{L}$ of logit vector and the respective encoded label is given as Equations \ref{eq::eq2} and \ref{eq::eq3}.
\begin{equation}
\mathcal{L}(\textbf{Y}) = - \sum_{i}^{E} \sum_{j}^{C} L_{ij} \textnormal{log} ( \textnormal{softmax}(\textbf{Y})_{ij}),
\label{eq::eq2}
\end{equation}

\begin{equation}
\textnormal{softmax} ( \textbf{Y} )_{ij} = \frac{\textnormal{exp}^{\textbf{Y}_{ij}}}{\sum_{j}^{C} \textnormal{exp}^{\textbf{Y}_{ij}}},
\label{eq::eq3}
\end{equation}
where $L$, $E$ and $C$ denote class labels, the number of training samples in $\textbf{Y}$ and the number of classes, respectively.

In summary, we have devised this network using a densely connected approach to better extract feature representations and strengthen the network's feature activation for predicting potential e-cheating.

\begin{table}[!t]
\renewcommand{\arraystretch}{1.3}
\caption{The configurations of each layer for the proposed network. $f$ refers to the size of feature maps and $k$ refers to the size of filter.}
\vspace{-5pt}
\label{tab::tab1}
\centering
\begin{tabular}{lc}
\toprule
\textbf{Network Layer} & \textbf{Configuration} \\ 
\midrule
$conv \:\: 3 \times 3$ & \textit{f}: 64$@$1$\times$23; \textit{k}: 1$\times$3; stride 2 \\
$LSTM \:\: Block_{1}$ & $\begin{bmatrix}
LSTM \:\: Cell\\ 
conv \:\: 2\times 2 
\end{bmatrix} \times 4$ \\
$transition$ & $conv$ 1$\times$1; $avgpool$; stride 2 \\
$LSTM \:\: Block_{2}$ & $\begin{bmatrix}
LSTM\:\: Cell\\ 
conv \:\: 2\times 2 
\end{bmatrix} \times 8$ \\
$LSTM \:\: Cell$ & 512$@$1$\times$6 \\
$flatten$ & 1$\times$6$\times$512 \\
\textbf{Y} & 1$\times C$ \\ 
\bottomrule
\end{tabular}
\vspace{-1.0em}
\end{table}

\section{PT Behavioural Database}
\label{sec::sec3}
Forming the test datasets for the e-cheating intelligent agent is the PT Behavioural Database, which we were responsible for its conceptualisation, design and data collection. This database consists of 663 candidate records acquired across different assessment periods for the "Introduction to Artificial Intelligence" course at Pyeongtaek University. All records were collected during the semesters of Spring 2020, Fall 2020 and Spring 2021 (totalling three semesters). The database is publicly accessible through \cite{ourdb2020}.

\subsection{Collection Setup}
\label{subsec::subsec31}
Data curation for the database was performed on the electronic class learning management system (LMS) that the university adopts. Assessment modules are important for the LMS system where it can be used to set up quizzes and assessment questions. Here, we collected seven datasets across three semesters. Each dataset is populated with 20 multi-choice questions that contain three levels of difficulty: `easy', `moderate' and `advanced'. During an assessment, the LMS automatically records candidates information such as their answers for each question, the scores for each question, the grades, the total time taken for completion (in minutes) and the IP addresses, which is shown in Table \ref{tab::tab2}.

\begin{table}[!t]
\renewcommand{\arraystretch}{1.3}
\caption{Several sample records from the Spring 2020 semester from our database. See \cite{ourdb2020} as complete version of data records for each semester.}
\vspace{-5pt}
\label{tab::tab2}
\centering
\begin{tabular}{cccccc}
\toprule
\multicolumn{1}{c}{\begin{tabular}[c]{@{}c@{}}\textbf{Q1}\\ \textbf{Score}\end{tabular}} & \multicolumn{1}{c}{\begin{tabular}[c]{@{}c@{}}\textbf{Q2}\\ \textbf{Score}\end{tabular}} & \textbf{$\cdots$} & \multicolumn{1}{c}{\begin{tabular}[c]{@{}c@{}}\textbf{Total}\\ \textbf{Score}\end{tabular}} & \multicolumn{1}{c}{\begin{tabular}[c]{@{}c@{}}\textbf{Time}\\ \textbf{(Minutes)}\end{tabular}} & \textbf{IP}\\ 
\midrule
5 & 2 & $\cdots$ & 94 & 13 & 175.116.139.44 \\
5 & 2 & $\cdots$ & 96 & 18 & 211.214.126.62 \\
5 & 2 & $\cdots$ & 82 & 16 & 180.71.78.211 \\ 
5 & 2 & $\cdots$ & 94 & 10 & 211.243.246.3 \\ 
5 & 2 & $\cdots$ & 94 & 10 & 211.243.246.3 \\ 
5 & 2 & $\cdots$ & 94 & 29 & 221.147.167.237 \\ 
5 & 2 & $\cdots$ & 76 & 11 & 121.165.205.164 \\ 
\bottomrule
\end{tabular}
\end{table}

\subsection{Training and Testing Protocols}
\label{subsec::subsec32}
We selected 95 records from the Spring 2020 semester to form the training protocol. These records were ensured that they do not overlap or contain duplicates in other datasets. To design the protocol for model development and training, we divided the dataset for training and cross-validations using a ratio of 80:20. Due to the imbalanced nature of the dataset, we further apply a data augmentation method to generate additional samples to represent cases of `suspected' behaviours. 
Next, we assigned the remaining datasets as test sets. In this testing scheme, the candidates' behaviours were determined based on their manner of answering assessment questions.

\subsection{Data Representation} 
\begin{figure}[!t]
\centering
\includegraphics[width=0.48\textwidth]{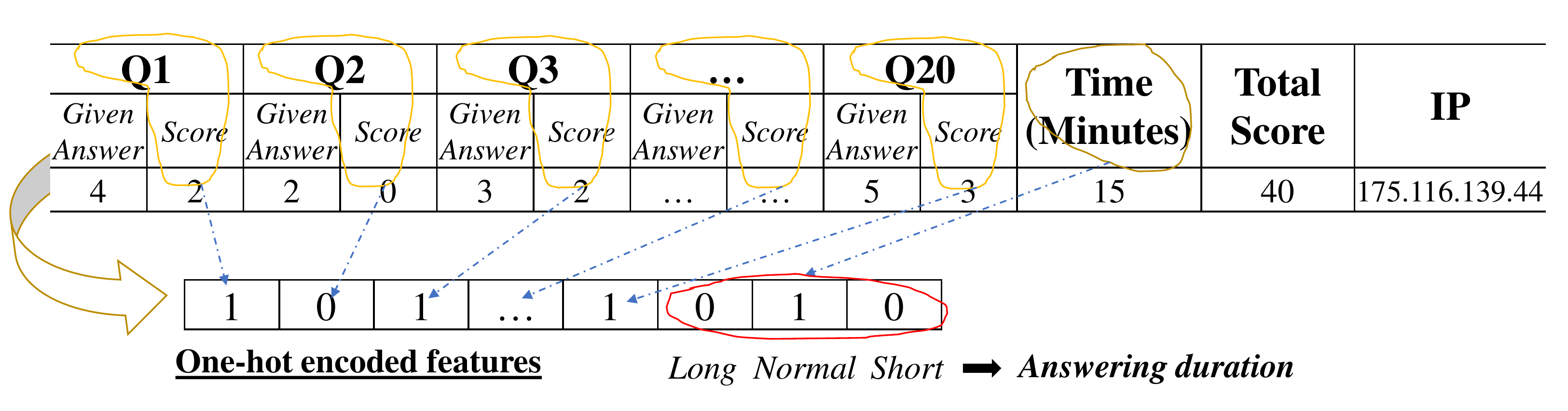}
\vspace{-10pt}
\caption{Generation of the one-hot encoded feature.}
\label{fig::fig3}
\end{figure}

We use one-hot encoding to transform each raw data record into binary features before training the behavioural monitor, which defines an assessment candidate's behaviour. 
Each record entry contains a candidate's answer to 20 multiple-choice questions, total assessment duration (in minutes), and final score for a sample assessment. We define the encoded input feature as $R \in 1 \times N$, where $N=23$. The first 20 elements represent the candidate's answers to the 20 questions $[1, 1, 1, 1, 1, 0, 1, 0, \cdots, 0]$ where the values 1 and 0 denoting a correct or incorrect answer, respectively. The last three elements define the answering duration, e.g, long, normal or short. Fig. \ref{fig::fig3} summarises the data representation, showing the processing of raw data into one-hot encoded features.

Each candidate's behaviour is classified into either of two states: `normal' or `suspected'. To identify these behaviours, we followed the concept by \cite{Ullah2018} to analyse behaviours. In defining a `suspected' behaviour, we assessed the answering duration where candidates have answered 90\% of the questions correctly according to the one-hot encoded features. If this speed was found to be too fast or too slow, they are labelled as `suspected'; the rest of the samples would be considered `normal'.

Two factors were considered when defining the speed of answering: the number of questions answered and the difficulty level for each question. Easier questions are answered more quickly. For example, we observed that when the level of difficulty for a question was set to `easy', most of the candidates could answer it within 10 to 20 seconds; if it was set to `moderate' or `advanced', they would require 30 to 40 seconds or 1--2 minutes, respectively. Note that the specifics of such labelling criteria depend on the assessment subjects or the evaluated courses.

\section{Experiments and Results}
\label{sec::sec4}
The relative performance of our system and network was evaluated across several experiments, with comparisons drawn from other similar benchmarks.
The configurations used for the networks are described in Section \ref{subsec::subsec41} and the experimental results are presented in Section \ref{subsec::subsec42}.

\subsection{Experimental Setup}
\label{subsec::subsec41}

\subsubsection{Configuration of DenseLSTM}
Our DenseLSTM network was implemented on the TensorFlow \cite{TensorFlow} platform. To configure, we applied a learning rate of $1.0\times 10^{-5}$ and an AdamOptimizer \cite{Kingma2015}, with the weight decay and momentum set to $1.0\times 10^{-4}$ and 0.9, respectively. For the experiments, the batch size was set to 16 and training was carried out across 250 epochs. Training was conducted using our PT Behavioural Database according to the protocols described in Section \ref{subsec::subsec42}. Processing was handled mainly by an Nvidia GeForce RTX 2080 Ti graphics processing unit.

\subsubsection{Configuration of Benchmark Networks}
\label{sec::confbn}
We selected some deep networks to evaluate the behavioural monitor's performance and compare them against the DenseLSTM. These are the Deep Neural Network (DNN) \cite{Kriegeskorte2019}, the LSTM \cite{Greff2017} and the Recurrent Neural Network (RNN) \cite{Rumelhart1986}, all of which have been successfully applied in past behavioural studies \cite{Zerkouk2019} and \cite{Hao2019}. When conducting the experiments, LSTM and RNN were implemented and optimised from scratch to our best attempts following the recommendations of \cite{Greff2017} and \cite{Rumelhart1986}. Training was similarly conducted using the PT Behavioural Database, following the protocols set in Section \ref{subsec::subsec42}. All training procedures were likewise performed on the RTX 2080 Ti.

\subsection{Experimental Results}
\label{subsec::subsec42}
\subsubsection{IP Detector Results}
\begin{figure*}[!t]
\centering
\subfloat[Test 1 Spring 2020]{\includegraphics[width=2.in]{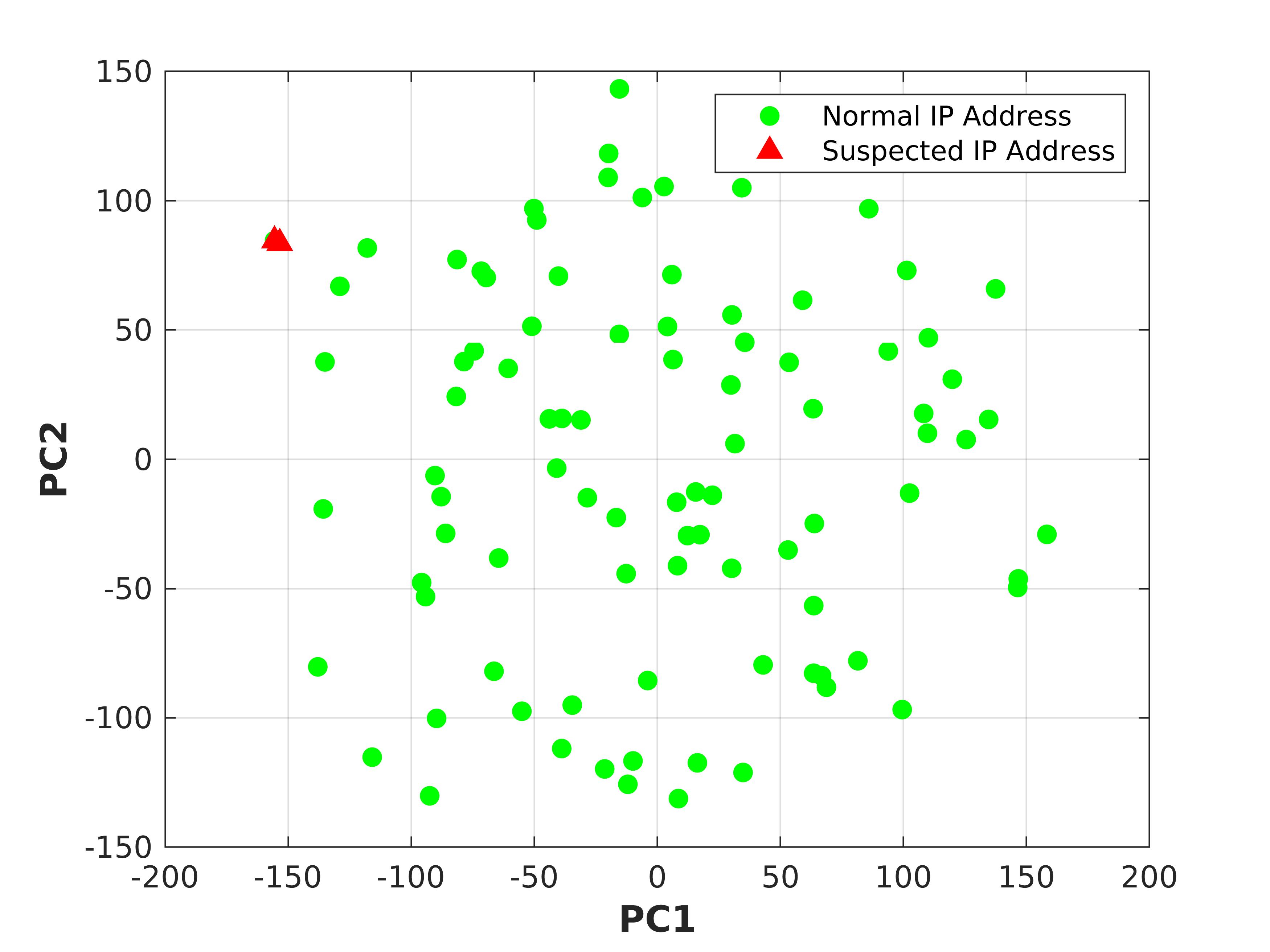}%
\label{fig::fig4a}}
\hfil
\subfloat[Test 2 Spring 2020]{\includegraphics[width=2.in]{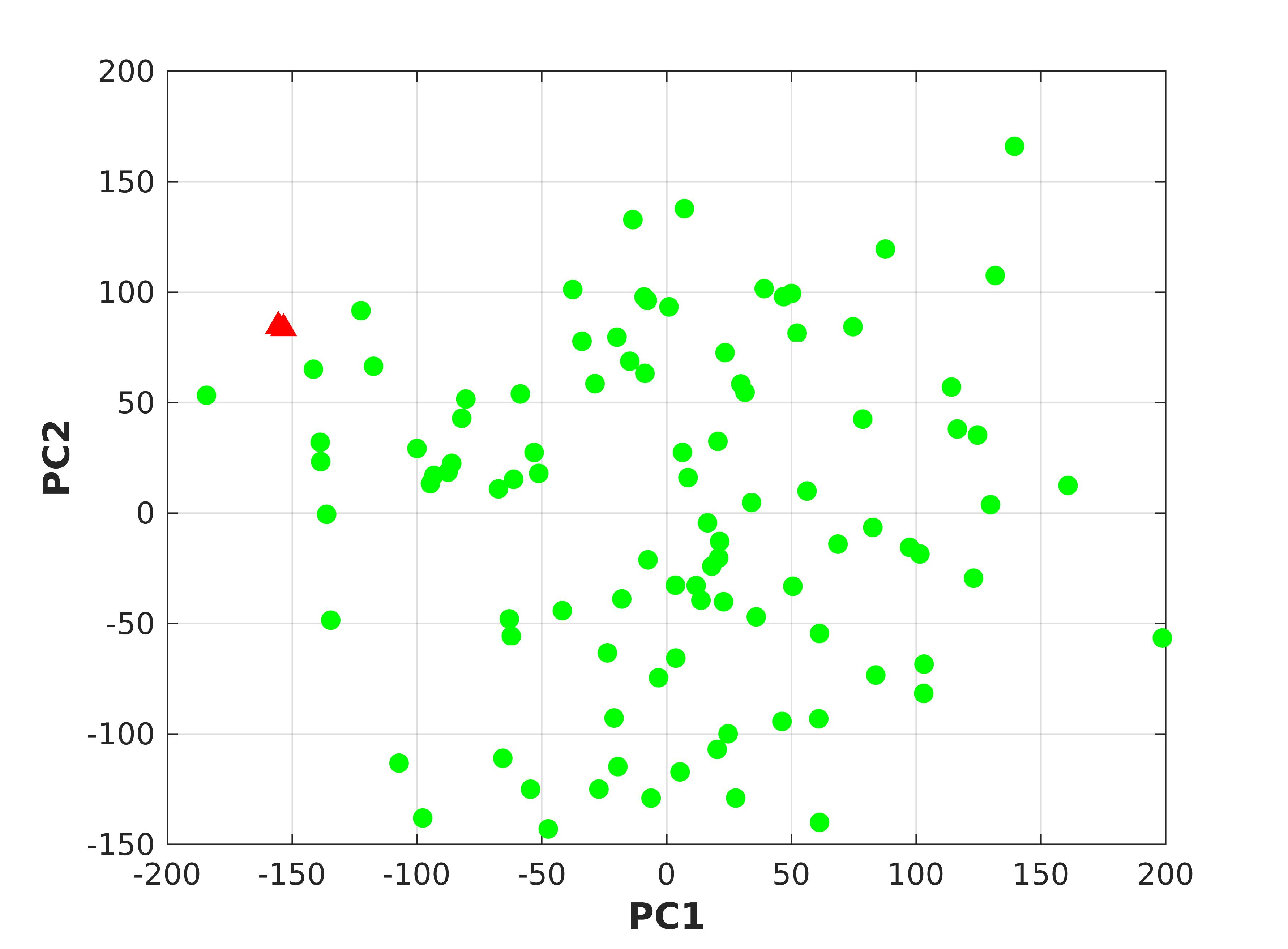}%
\label{fig::fig4b}}
\hfil
\subfloat[Test 3 Fall 2020]{\includegraphics[width=2.in]{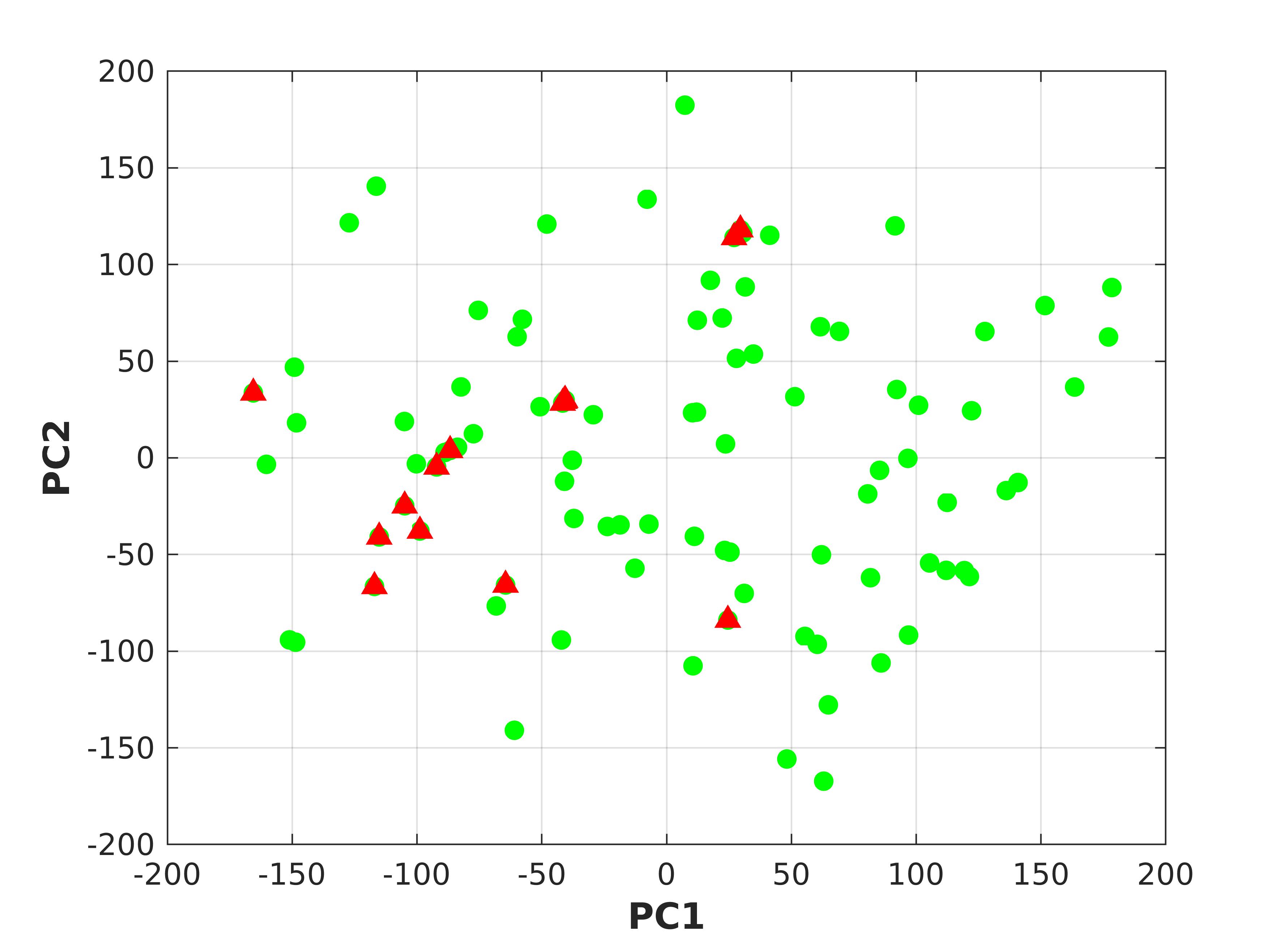}%
\label{fig::fig4c}}
\hfil
\subfloat[Test 4 Fall 2020]{\includegraphics[width=2.in]{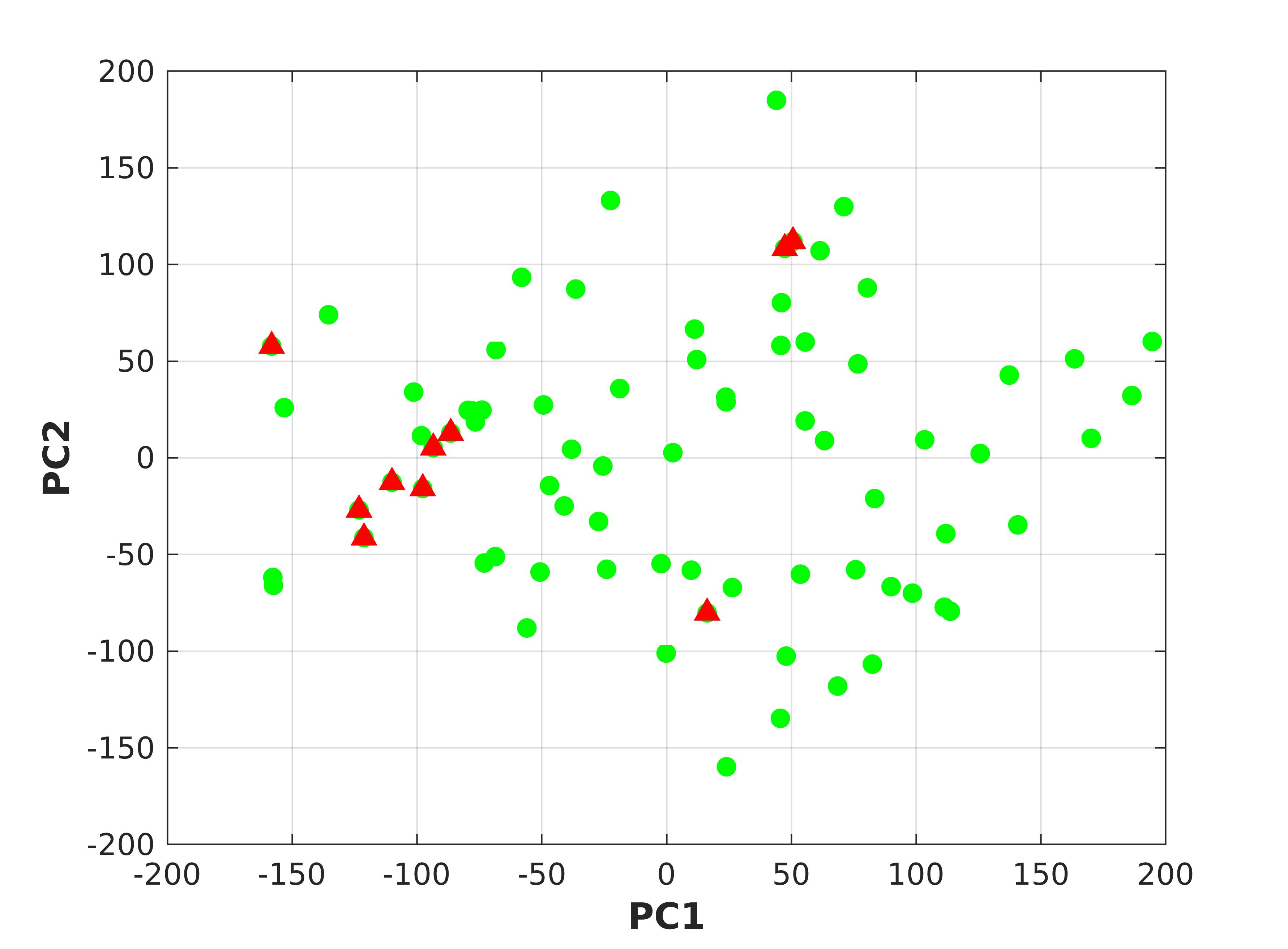}%
\label{fig::fig4d}}
\hfil
\subfloat[Test 5 Spring 2021]{\includegraphics[width=2.in]{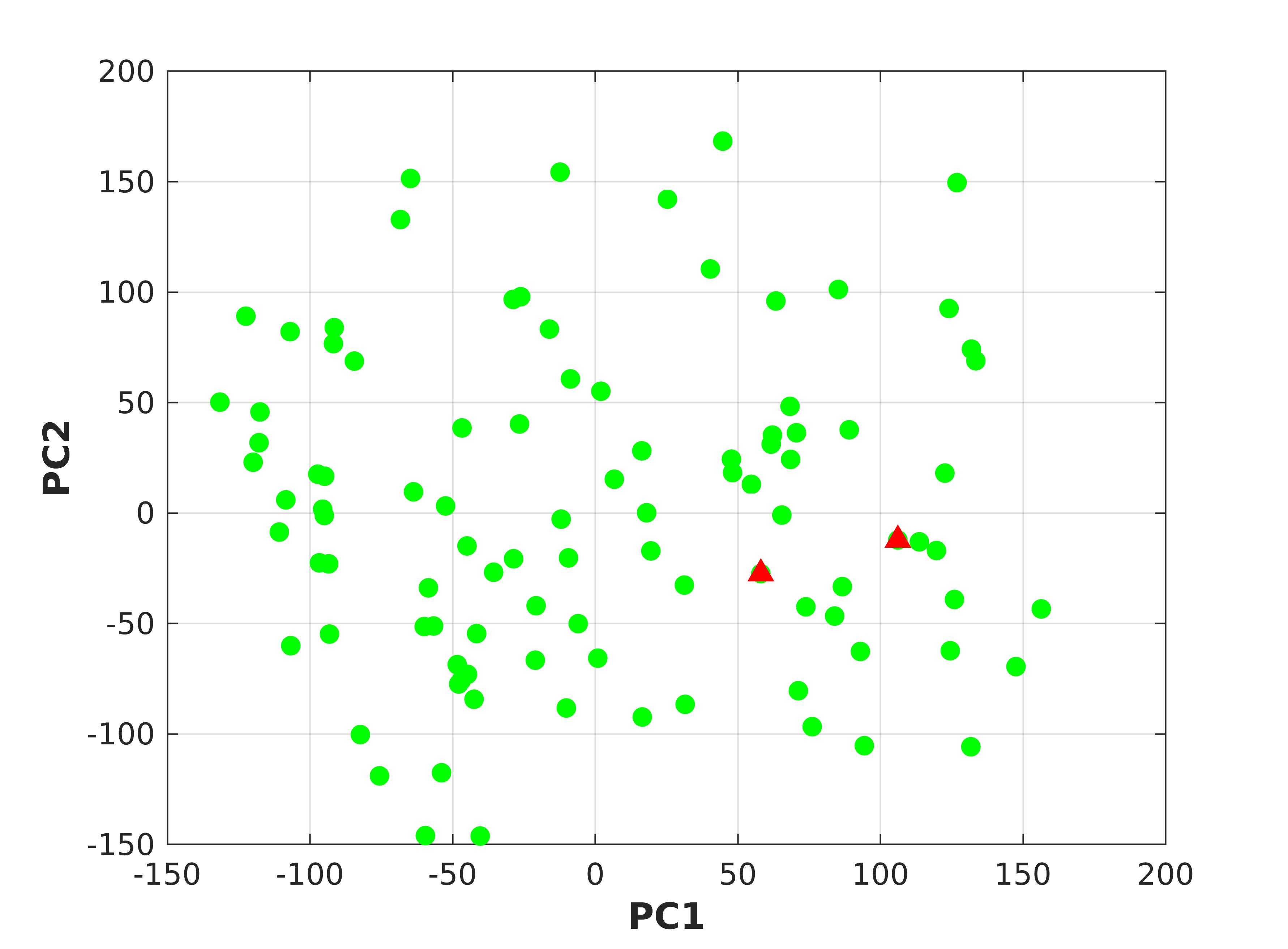}%
\label{fig::fig4e}}
\hfil
\subfloat[Test 6 Spring 2021]{\includegraphics[width=2.in]{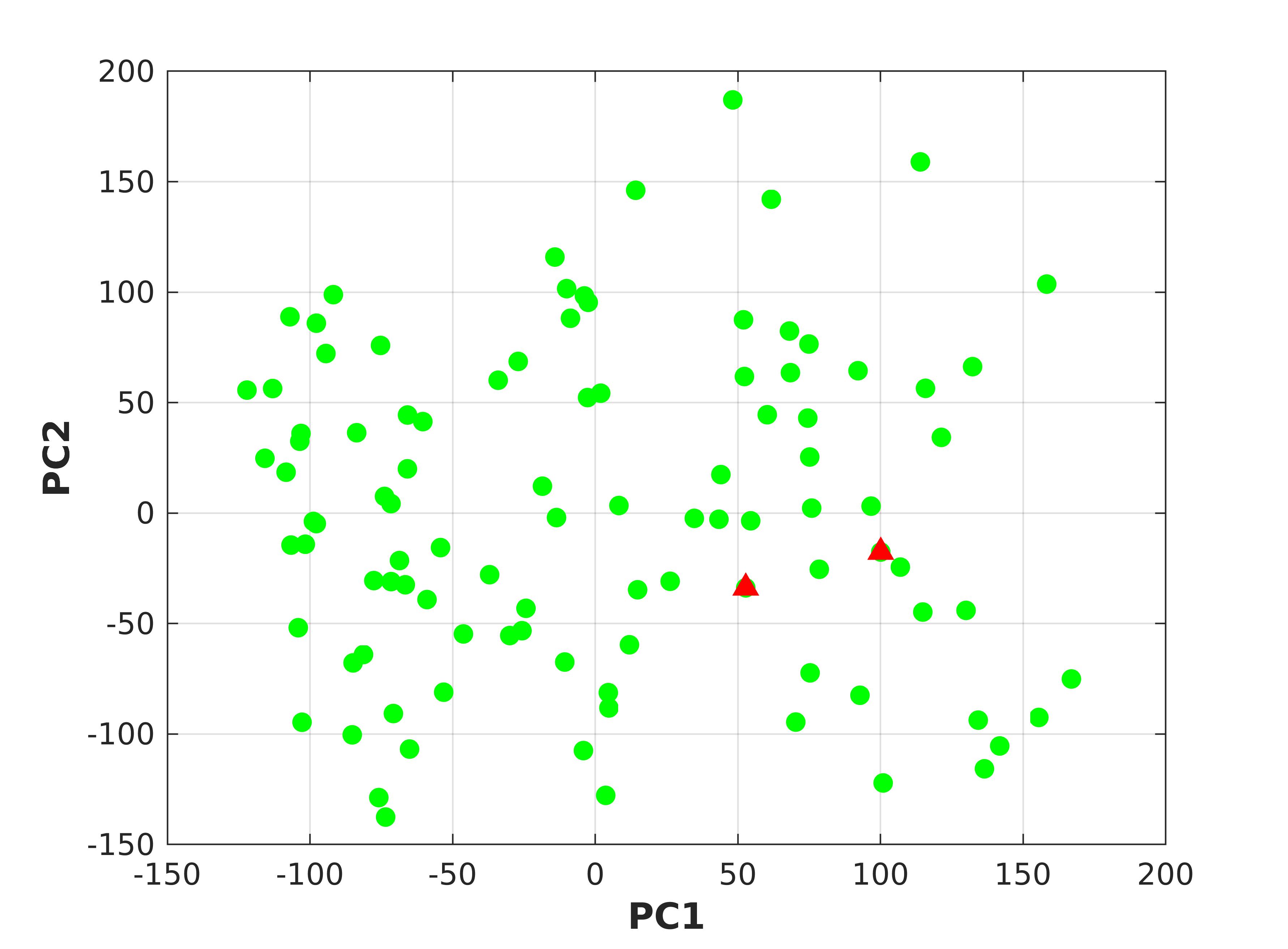}%
\label{fig::fig4f}}
\caption{Performances of the network IP detection agent for suspect IP addresses on six test sets.}
\label{fig::fig4}
\end{figure*}

We evaluated the performance of the IP detection agent across six test sets according the protocol outlined in Section \ref{subsec::subsec41}. Principal component analysis was used to visualise `normal' and `suspected' IP addresses in the test sets shown in Fig. \ref{fig::fig4}. From Fig. \ref{fig::fig4}, there are less than 10 cases where IP addresses were flagged as `suspected' in a given test set. We observed that many `suspected' IP addresses originate from geographically similar locations, i.e., the candidates were roommates, housemates or located within the same premise.
Results shown the IP detector has provided a foundation for security analysis by estimating the candidates' locations during online assessments. In addition, this agent can easily filter potentially fraudulent activities as an early precaution.

\subsubsection{Behavioural Monitor Results}
To objectively evaluate the performance of our DenseLSTM network in a practical setting, we applied the experiments across six test sets and compared the results against those obtained from the benchmark networks described in Section \ref{sec::confbn}. From Table \ref{tab::tab3}, of the different networks tested, the DenseLSTM network achieved the highest accuracy (90.72\%) for overall performance, followed by the LSTM network with an accuracy of 88.22\%. Our network outperformed existing benchmark models by 2.5\%, demonstrating its capabilities against other benchmark approaches.

Through the identification of system improvements, we observed that the accuracy of our network was higher than 90\% when analysing the behaviour of candidates for the majority of test sets, with values of 97.77\% for Test 1, 91.92\% for Test 2, 91.96\% for Test 3 and 90.18\% for Test 5, respectively (as shown in Table \ref{tab::tab3}). The LSTM network, which was the second-best, demonstrated comparable accuracy scores of 88.22\% overall. In contrast, the results of DNN was lower, with an accuracy score of 65.74\% overall. In addition, our investigations of DNN revealed error rates of 12--14\% overall, which was associated with several false preventions that occurred as a result of some candidates that answered very slowly.

\begin{figure*}[!t]
\centering
\subfloat[Test 1 Spring 2020]{\includegraphics[width=2.in]{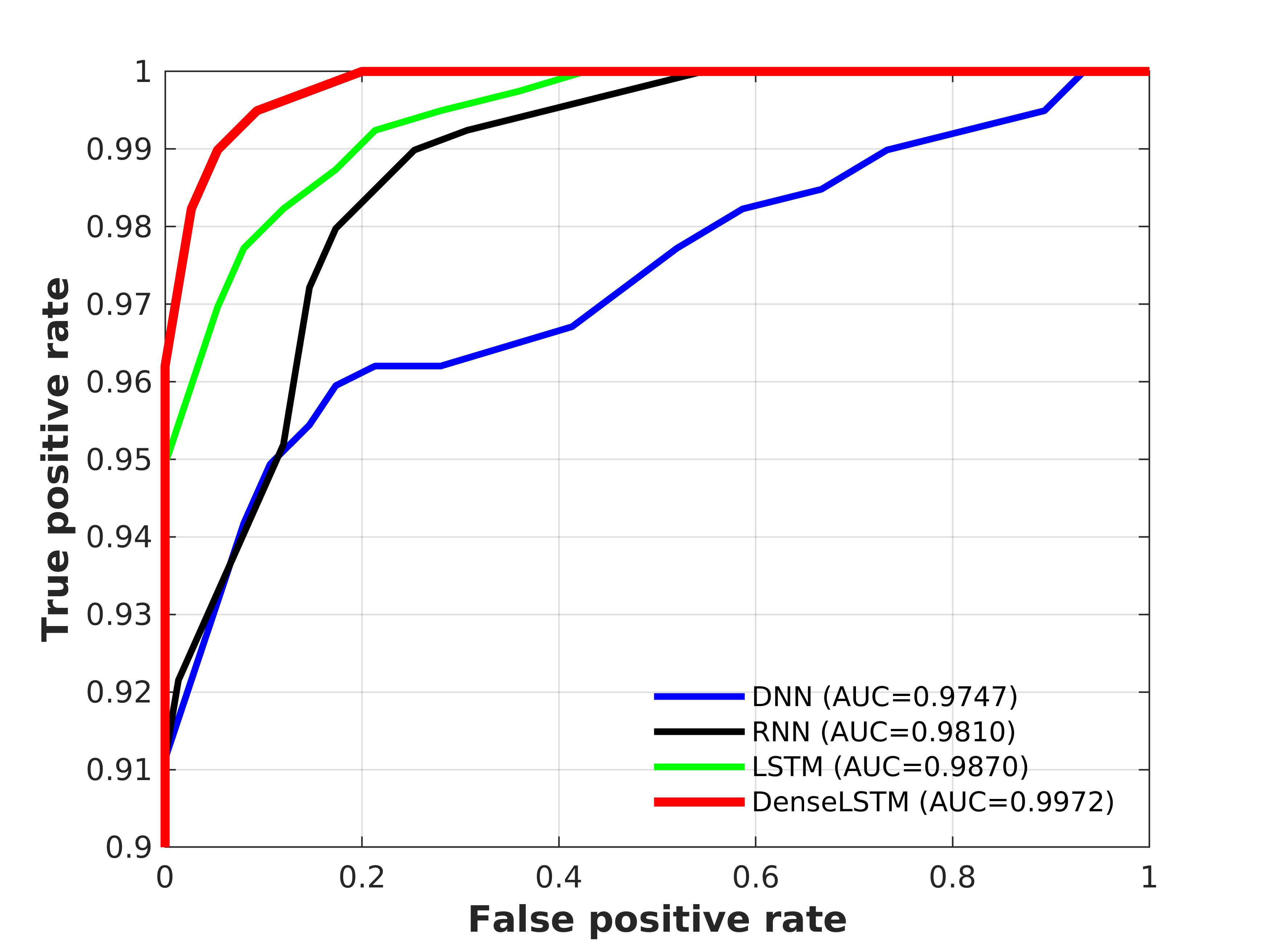}%
\label{fig::fig5a}}
\hfil
\subfloat[Test 2 Spring 2020]{\includegraphics[width=2.in]{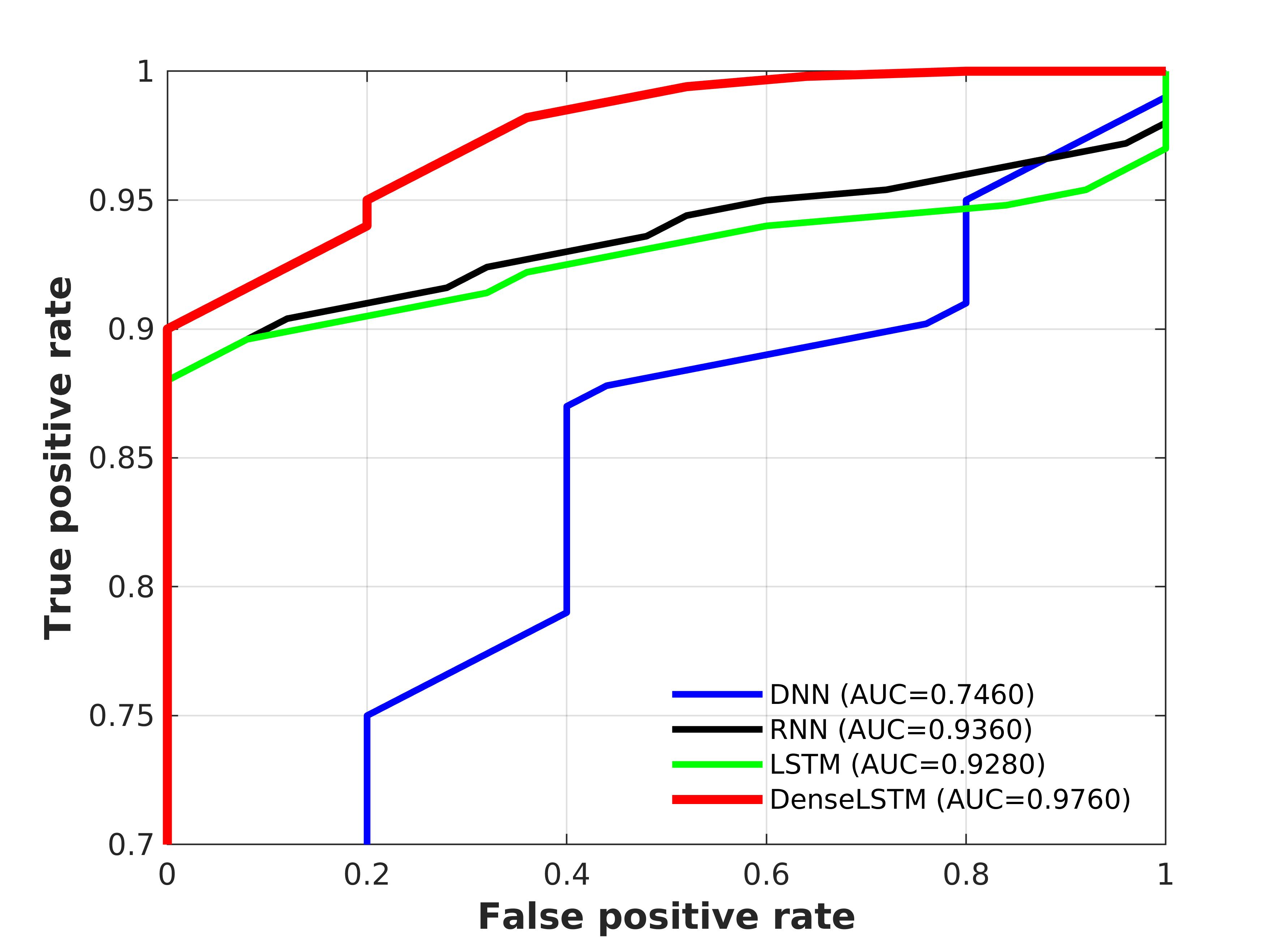}%
\label{fig::fig5b}}
\hfil
\subfloat[Test 3 Fall 2020]{\includegraphics[width=2.in]{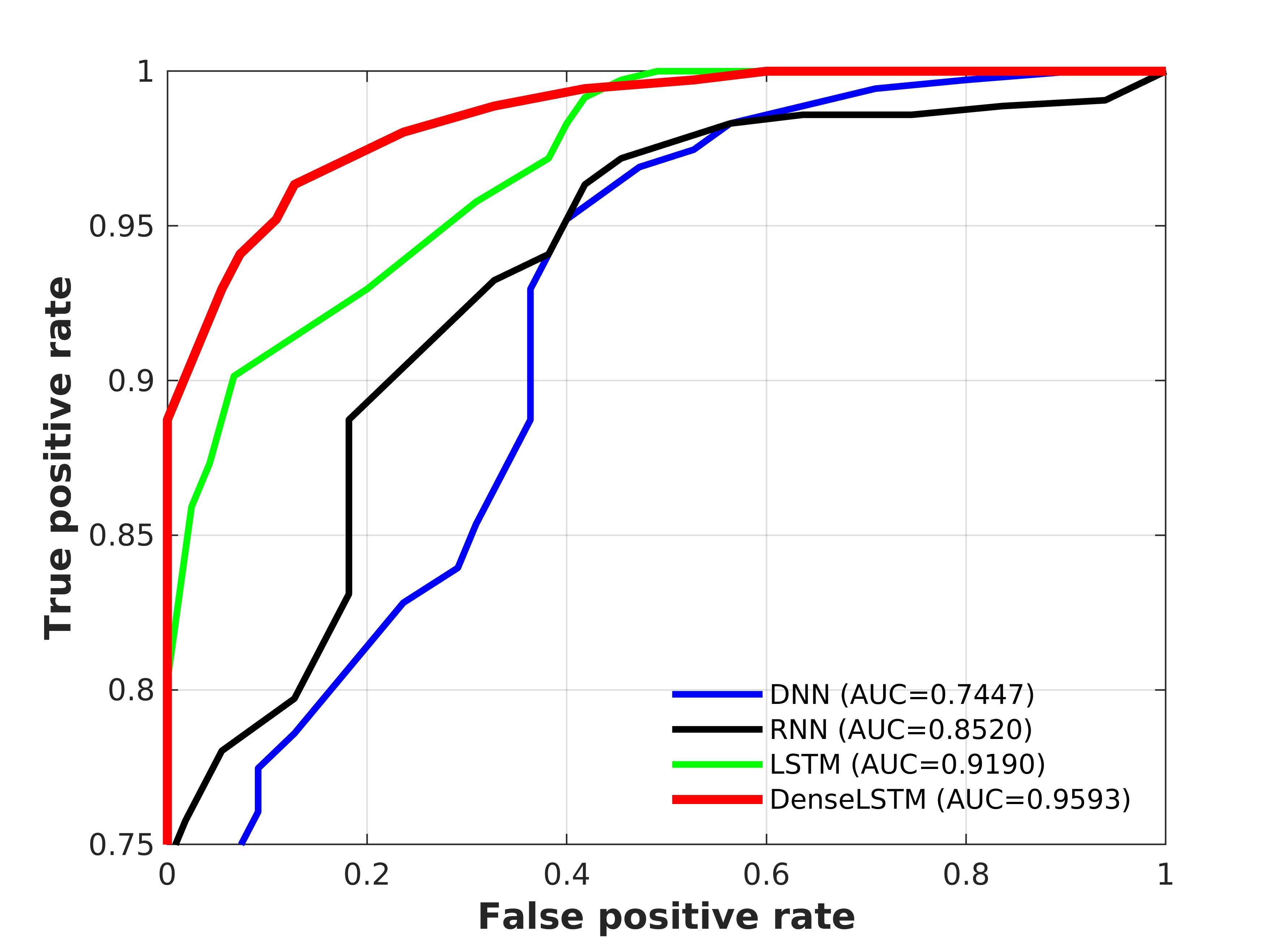}%
\label{fig::fig5c}}
\hfil
\subfloat[Test 4 Fall 2020]{\includegraphics[width=2.in]{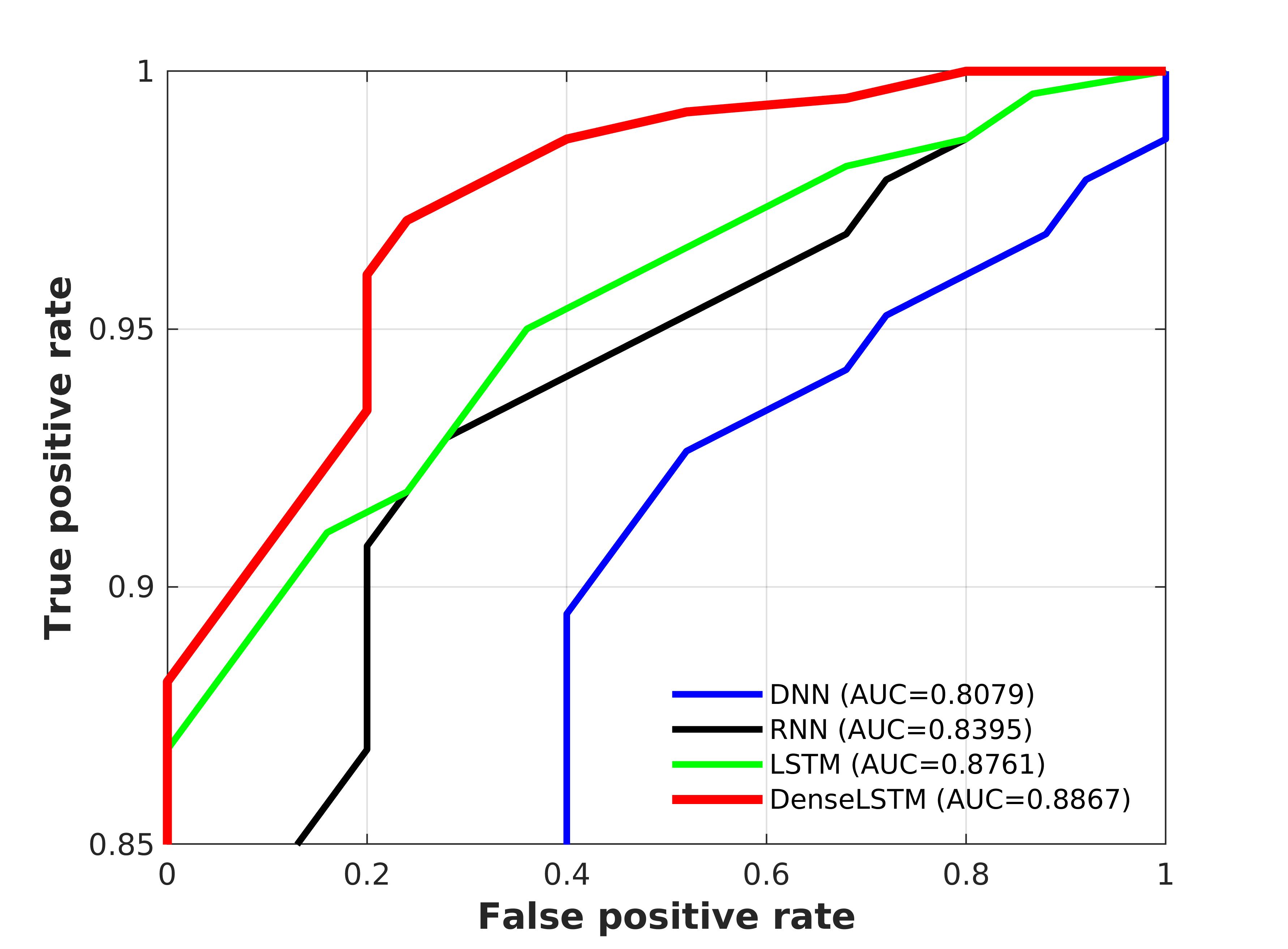}%
\label{fig::fig5d}}
\hfil
\subfloat[Test 5 Spring 2021]{\includegraphics[width=2.in]{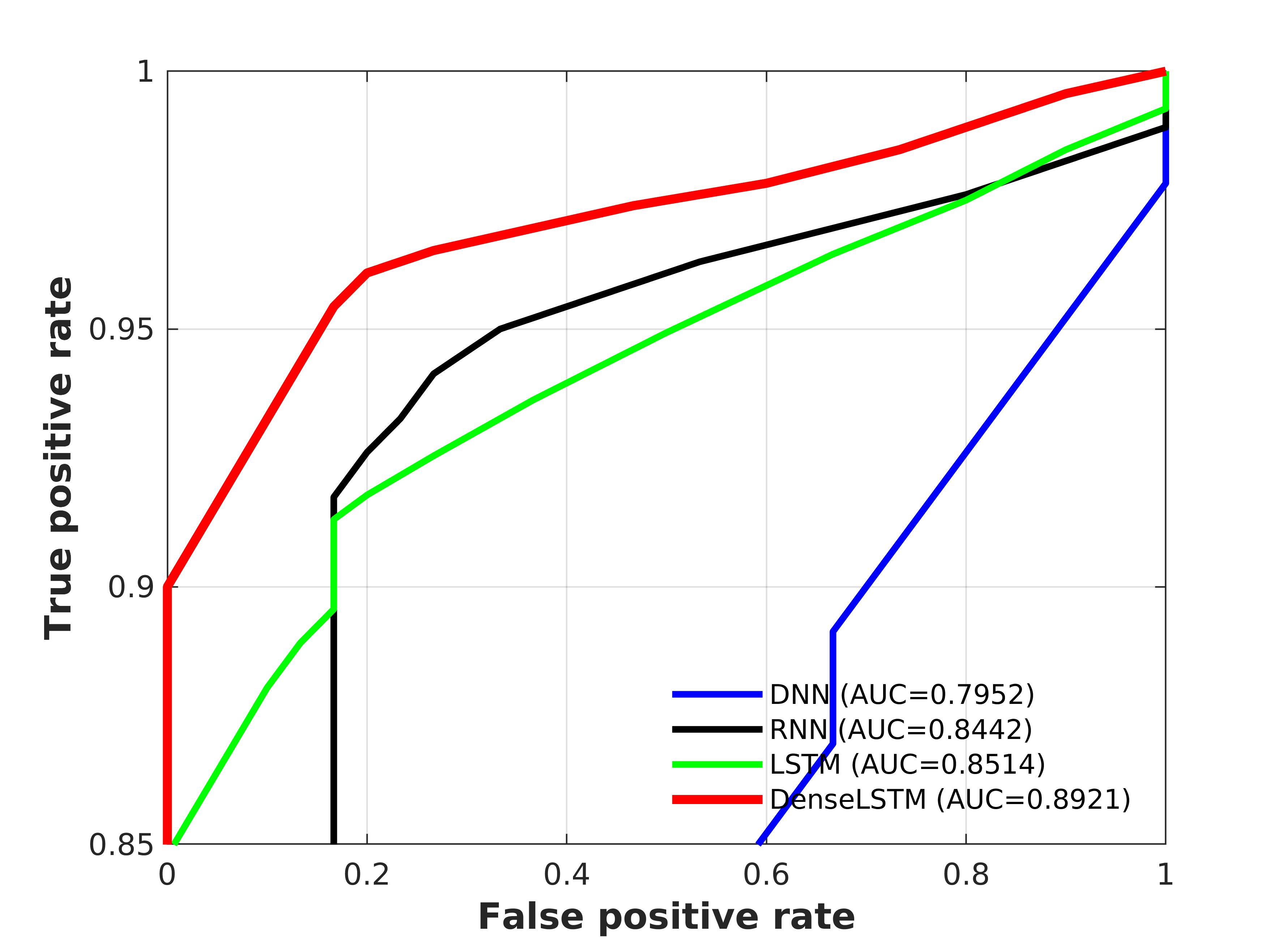}%
\label{fig::fig5e}}
\hfil
\subfloat[Test 6 Spring 2021]{\includegraphics[width=2.in]{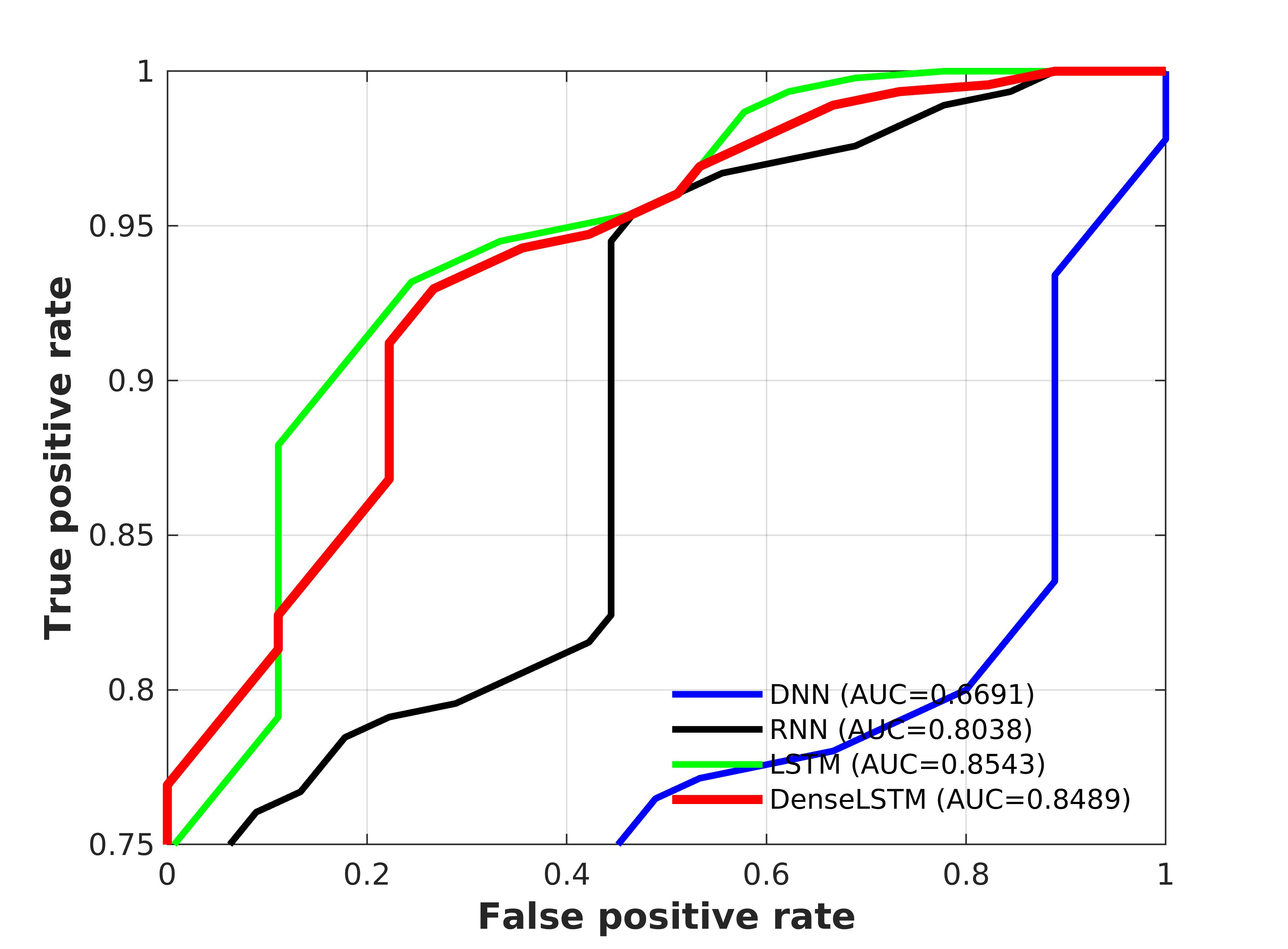}%
\label{fig::fig5f}}
\caption{Performance comparisons of the ROC curve for six test sets.}
\label{fig::fig5}
\end{figure*}

To further refine our network, we focused on characterising its sensitivity and specificity by applying the Receiver Operating Characteristics (ROC) and Area Under the ROC curve (AUC) parameters. This characterisation summarises the trade-off between the true and false-positive rates using different probability thresholds. In Fig. \ref{fig::fig5}, DenseLSTM achieved the highest AUC values of 0.9972, 0.9760, 0.9593, 0.9043, 0.8867 and 0.8921 for Test 1 to Test 5, respectively, excluding Test 6. These comparisons demonstrate our network's ability to outperform most other benchmark networks in classifying candidates' `normal' and `suspected' behaviours at online assessments with low sensitivity and high specificity.

\begin{table}[!t]
\renewcommand{\arraystretch}{1.2}
\caption{Performance evaluation of online assessments for the Spring 2020, Fall 2020 and Spring 2021 semesters. Figures of highest accuracy are highlighted in bold.}
\vspace{-5pt}
\label{tab::tab3}
\centering
\begin{tabular}{lcccc}
\toprule
\textbf{Testset} & \textbf{DNN} & \textbf{RNN} & \textbf{LSTM} & \textbf{DenseLSTM} \\ 
\midrule
Test 1 Spring, 2020 & 82.74\% & 87.20\% & 94.49\% & \textbf{97.77\%} \\
Test 2 Spring, 2020 & 52.68\% & 85.02\% & 89.29\% & \textbf{91.92\%} \\
Test 3 Fall, 2020 & 56.65\% & 90.12\% & 89.29\% & \textbf{91.96\%} \\
Test 4 Fall, 2020 & 72.92\% & 85.42\% & 87.50\% & \textbf{88.54\%} \\
Test 5 Spring, 2021 & 64.29\% & 85.71\% & 89.29\% & \textbf{90.18\%} \\
Test 6 Spring, 2021 & 65.18\% & 80.35\% & 79.46\% & \textbf{83.93\%} \\
\midrule
\textbf{Overall} & 65.74\% & 85.64\% & 88.22\% & \textbf{90.72\%}\\
\bottomrule
\end{tabular}
\end{table}

\subsection{Discussion}
Experimental analysis and results show that our proposed approach can successfully address challenges on online assessment misconducts, preventing candidates `suspected behaviour. Further justifying our observation, there was a plagiarism case where a pair of twins who had sat for an online assessment with the `suspected' IP address 211.243.246.3 (shown in Table \ref{tab::tab2}). Although the IP detector had assigned unique sets of assessments, the behavioural monitor can flag them as `suspected' in response to their answering behaviours. After an investigation by the lecturer, it was found that the candidates had similar answering durations, including the time needed to complete the assessment. Results indicate that our system enables course coordinators to monitor candidates through network protocols and behavioural analysis to identify potential misconduct.

With the experiments conducted across six test sets, results have shown consistently high accuracy in favour of our system. They validate the performance of our proposed network to be superior in detecting `suspected' behaviours at online assessments due to its ability to better maintain a ``collective knowledge'' from the extracted features within the subsequent layers of the network, as opposed to simply ``forgetting'' useful information. This supports our assumption that the DenseLSTM performs better than other neural networks for this application.

\section{Conclusion}
Online learning is a new and exciting opportunity gaining momentum for both students and academic institutions alike, presenting unique opportunities and challenges. The primary concern of online assessments is academic dishonesty and misconducts in the form of cheating and plagiarism, whereby candidates attempt to achieve through numerous, often creative, avenues. Therefore, it is the responsibility of academic institutions to implement effective measures to deter these misconducts. This paper discusses the concerns surrounding online assessment misconducts and offers plausible mechanisms of monitoring and curtailing such incidences using AI.

We have demonstrated the effectiveness of a proposed e-cheating intelligent agent, which successfully incorporates IP detector and behavioural monitor protocols. However, the presented results are limited in cases where the assessment is designed using multiple-choice questions, higher bandwidth 3G/4G networks, etc. The current state of online assessments can occur everywhere through other delivery forms, such as essay formats, 2G network and other environmental factors. As future work, we plan to continue the development of this system to achieve testings across various courses in unconstrained environments, which we hope will facilitate the practical usage of detecting similarly abnormal behaviour in candidates during online assessments.




\IEEEtriggeratref{13}


\bibliographystyle{IEEEtran}
\bibliography{IEEEabrv, myref}
%



\end{document}